\documentclass{blois}

\def\be{\begin{equation}}
\def\ee{\end{equation}}
\def\bea{\begin{eqnarray}}
\def\eea{\end{eqnarray}}

\newcommand{\Photo}{}

\begin{document}
\vspace*{4cm}
\title{Primordial black holes}

\author{Guillermo Ballesteros}

\address{Departamento de F\'isica Te\'orica, Universidad Aut\'onoma de Madrid (UAM),\\
Campus de Cantoblanco, 28049 Madrid, Spain\\ 
Instituto de F\'isica Te\'orica (IFT) UAM-CSIC, Campus de Cantoblanco, 28049 Madrid, Spain}

\maketitle\abstracts{
The possibility that dark matter could be primordial black holes is discussed with an emphasis on the most commonly studied inflationary dynamics that could have produced them. 
 }
\section{Introduction}
Primordial black holes (PBH) \cite{Zeldovich:1967lct,Hawking:1971ei} are hypothetical black holes (BH) that are  assumed to have formed before the period of big bang nucleosynthesis. Much  of the present motivation to study them comes from the possibility that they could constitute the dark matter of the Universe. Current data, combined with our understanding of the physics of these objects, indicates that PBH in the mass range that spans approximately from $10^{-16}$~M$_\odot$ to $10^{-12}$~M$_\odot$ can account for all the dark matter --see \cite{Green:2020jor} for a review of recent bounds-- and have Schwarzschild radii between $10^{-3}$~\r{A} and 0.1~{\r{A}.\footnote{The Schwarzschild radius of the Sun is about  3~km and the ``diameter'' of a Hydrogen atom is $\sim 1$~\r{A}.} PBH outside this asteroid-mass \footnote{The mass of the large asteroid (or dwarf planet) Ceres is near the upper limit of this range.}  range (or window)  may contribute to a small percentage of the dark matter. The lower end of the window comes from the non-observation of $\gamma$  and X-rays from PBH evaporation that would occur via Hawking radiation, see \cite{Carr:2009jm} and also \cite{Ballesteros:2019exr,Iguaz:2021irx}. The upper limit comes instead from micro-lensing observations of celestial bodies, specifically with the Hyper Supreme-Cam of the telescope Subaru \cite{Niikura:2017zjd}.\footnote{See \cite{Mroz:2024wia} as well.}

In 2019 there was no consensus regarding the viability of the asteroid-mass window to account for all the dark matter. It had been argued that the microlensing bounds existing at the time, together with limits from femtolensing of $\gamma$-ray bursts and the possible encounters of PBH with neutron stars and white dwarf stars \cite{Graham:2015apa} constrained the asteroid-mass range in such a way it could only account for $\sim 10\%$ of the dark matter. By 2020 the window had opened, after those limits were reassessed \cite{Katz:2018zrn,Montero-Camacho:2019jte}.\footnote{See also \cite{Sugiyama:2019dgt}.} Today, the asteroid-mass range is widely considered the most promising mass region to explain the dark matter with PBH. Narrowing this window any further is not an easy task, due to intrinsic experimental limitations. Interesting progress has been done on the lower mass limit by considering not the diffuse (extragalactic) $\gamma$-ray background but emission from our galaxy, see e.g.~\cite{Laha:2019ssq}. However, these bounds do not reduce the overall width of the window very significantly on a qualitative level. Most bounds are computed assuming a single mass for the PBH. Other assumptions, commonly referred as broad mass functions, have been argued to narrow the width of the asteroid-mass window for PBH dark matter \cite{Carr:2017jsz}. Finding novel experimental or theoretical ways of constraining the PBH asteroid-mass range for dark matter (or potentially leading to a discovery) is an endeavor worth pursuing. 

The interest in PBH as a dark matter candidate surged in 2016 after the announcement of the first detection of the gravitational waves emitted by a binary BH 
merger, which was named GW150914~\cite{LIGOScientific:2016aoc}. The individual  BH masses of the binary inferred from the waveform of that event are  $\sim 30$~M$_\odot$. The question of whether BH more massive than our Sun by an order of magnitude or two could be the dark matter was then swiftly put forward \cite{Bird:2016dcv}. However, it was soon pointed out that a comparison between the merger rate inferred from that single event and theoretical expectations assuming binary formation in the early Universe \cite{Nakamura:1997sm} --the largest contribution known to the  merger rate-- only allow a sub-percent fraction of the dark matter to be in  the form of PBH in that mass range \cite{Sasaki:2016jop}. Recent analyses which include $\mathcal{O}(50)$ mergers have confirmed this result; see for instance \cite{Franciolini:2021tla}. It has been suggested that PBH may form clusters that would have an impact on their merger rate, see e.g.\ \cite{Jedamzik:2020ypm}. However, there are arguments that indicate that the importance of clustering may not be very significant in the PBH mass range relevant for LIGO for commonly studied PBH formation scenarios, see for instance \cite{Ballesteros:2018swv,Crescimbeni:2025ywm}. There are also other tight limits on the abundance of PBH in that range, coming from different physics, see \cite{Green:2020jor}.

It is worth stressing that regardless of whether PBH are the main component of the dark matter, discovering a PBH (e.g.\ via the gravitational waves emitted from the merger of a binary) would be momentous. A BH with mass well below M$_\odot$ would be most certainly primordial. 

PBH have been considered as possible sources for particle creation within and beyond the Standard Model (including dark matter). They have been proposed as well to be the seeds of supermassive BH. Various potential mechanisms have been put forward to explain how PBH could have formed: the collapse of large radiation or matter overdensities seeded by inflation or reheating~\footnote{See however \cite{Ballesteros:2024hhq}.}, phase transitions, topological defects, false vacuum decay and others. For a non-exhaustive list of these topics see \cite{Carr:2020gox}. The first of these ideas for PBH formation is discussed below.

\section{Primordial black holes from inflation}
The allure of the idea that PBH of inflationary origin could explain the dark matter comes, in essence, from its apparent simplicity. Despite its caveats, primordial inflation is the most compelling and popular mechanism that has been proposed to set the initial conditions determining the Cosmic Microwave Background (CMB) temperature anisotropies and the distribution of matter at large scales. Perhaps inflation also seeded large inhomogeneities leading to the formation of PBH abundant enough to account for all the dark matter. These PBH could have formed solely from Standard Model particles. The evidence for dark matter has been a most significant reason to entertain a variety of ideas for physics beyond the Standard Model \cite{Cirelli:2024ssz}. If PBH are the dark matter, we should be looking for models of the very early Universe allowing their formation, and for methods to test them. 

The PBH dark matter hypothesis leads to the question of what kind of time evolution can produce dense enough regions that can collapse into BH under their own gravitational pull. In the most commonly considered scenario, entire Hubble patches collapse during radiation domination to form PBH. The problem of estimating the mass and abundance of these PBH is framed in terms of the statistics of the fluctuations describing density inhomogeneities in Fourier space during radiation domination. A collapsing Hubble patch forms a PBH whose mass is determined by the volume of the patch at the time the collapse starts. This occurs when the inverse of the comoving Hubble radius of the patch becomes comparable to a characteristic comoving wave number, $k$, at which density fluctuations have a large variance (which enhances the probability of the collapse). A simple calculation shows that the relation between $k$ and the PBH mass $M$ is $M(k) \simeq 10^{-14}(10^{13}\, k^{-1}\, {\rm Mpc}^{-1} )^2\, {\rm M}_{\odot}$. The well-known Press-Schechter formalism, originally intended for halo formation, has been often applied to obtain quantitative estimates of the abundance.  This is computed integrating the probability distribution function of density fluctuations from some threshold up to infinity. The threshold required for collapse can be estimated using numerical methods, and a value that has been commonly used is $\sim 0.45$.  

Inflation enters into the game, as anticipated, determining the statistical properties of the Hubble patches by setting the initial conditions for their evolution during radiation domination. Assuming a Gaussian distribution for the density fluctuations with comoving wave number $k$ in the radiation epoch and using a linear relation to relate them to inflationary perturbations, one finds that a power spectrum of primordial curvature perturbations $\mathcal{P_\zeta}$ of size $\sim 10^{-2}$ around $k$ is required to account for all the dark matter with PBH of mass $\sim M(k)$.  This means that $\mathcal{P_\zeta}$ has to grow by seven orders of magnitude in between 0.01~Mpc$^{-1}$ (a wave number representative of the CMB scales) and the relevant scale $k$. Taking $k\simeq 10^{13}$~Mpc$^{-1}$, right in the asteroid-mass window, this amounts to an average growth of $\Delta \log \mathcal{P_\zeta}\sim 1/2\, \Delta \log k$ between those scales. Such a large increase of $\mathcal{P_\zeta}$ does not occur in standard models of slow-roll inflation. A possible type of inflationary dynamics compatible with this enhancement will be discussed below. 

These approximations to obtain the masses and the abundance of PBH suffer from several uncertainties that have been studied. The assumption of Gaussian fluctuations overlooks the fact that the relation between radiation density and primordial inflationary fluctuations is not linear. Besides, the latter can be significantly non-Gaussian, and this is can be expected in models of inflation that break slow-roll. Importantly, the PBH abundance is sensitive to the tail of the probability distribution of $\zeta$. Usual perturbation theory is not an adequate tool to describe the regime of large $\zeta$ and different methods have been and are being developed and used to study this problem, see e.g. \cite{Ezquiaga:2019ftu,Figueroa:2020jkf,Pattison:2021oen,Celoria:2021vjw,Ballesteros:2024pwn,Caravano:2024moy,Ballesteros:2024pbe}. It has also been argued that the threshold for collapse is not a universal number, but rather a quantity that depends on the shape of $\mathcal{P_\zeta}$ as function of $k$ \cite{Germani:2018jgr}. In general it can be expected to depend on the statistical properties of the fluctuations. Besides, the Press-Schechter formalism may not be the most adequate tool to compute the abundance.  Other methods, such as peaks theory, have been put forward and shown to give different results. Finally, rather than the radiation density contrast, a related real-space variable called compaction function has been proposed to model the properties of the collapse. Notwithstanding these caveats, the simple approximations discussed above are often used as order of magnitude estimates for PBH masses and abundances. 

\section{Ultra slow-roll inflation}

The simplest models of inflation compatible with cosmological data consist of a single scalar~$\phi$ (the inflaton) rolling down a potential with small derivatives in Planck units. If the acceleration of $\phi$ is negligible, the amplitude of the spectrum of curvature perturbations is $\mathcal{P_\zeta}\sim (H/m_p)^2(H/\dot\phi)^2$, where $H$ is the expansion rate of the Universe and $\dot\phi$ the velocity of the inflaton. This suggests that a {\it very} slowly rolling scalar (with a {\it very} flat potential) can generate large amplitude curvature fluctuations. However, the potential cannot be too flat everywhere, for two reasons. First: a potential that is excessively flat would not be in agreement with the CMB. Second: inflation has to end eventually. These constraints allow for a potential that is very flat in a restricted range of field values, with larger slopes away from it. In 1994 a piece-wise potential with these features was proposed for PBH dark matter \cite{Ivanov:1994pa}. It extended an even simpler model of inflation proposed two years earlier \cite{Starobinsky:1992ts}. In models such as these (and in more sophisticated ones sharing the main key features) the inflaton may enter an ultra slow-roll regime~\cite{Leach:2000yw,Kinney:2005vj}, in which its acceleration balances the friction effect due to the curvature of the Universe: $\ddot\phi+3H\dot\phi\simeq 0$. This is possible when the derivative of the potential is negligible. In this regime, the aforementioned slow-roll approximation $\mathcal{P_\zeta}\propto 1/\dot\phi^2$ tends to be insufficient to describe the spectrum of primordial perturbations well, see~\cite{Ballesteros:2017fsr}. The reason is that a normally decaying solution to the (linear) equation of motion for the fluctuations becomes growing in ultra-slow roll, surpassing the one that dominates in slow-roll~\cite{Leach:2001zf}. 

In addition to a large enough $\mathcal{P_\zeta}$ at appropriate scales, inflationary models of PBH dark matter need to be in agreement with the CMB, sustain accelerated expansion encompassing enough scales to solve the horizon and flatness problems of the hot big-bang and allow for inflation to end so that the Universe can reheat. Two possible single-field models aiming to fulfill these requirements were proposed in \cite{Ballesteros:2017fsr}. See also \cite{Garcia-Bellido:2017mdw,Kannike:2017bxn}. The models of \cite{Ballesteros:2017fsr} feature potentials that grow approximately as~$\phi^4$ and a coupling of $\phi$ to the scalar curvature, $R$, of the form $\phi^2R$. Upon appropriate changes of variables, this setup leads to a potential (in the so-called Einstein frame) with a shape of a kind favored by the CMB~\cite{Planck:2018jri}. The two models differ on the specifics of the potential in the original Jordan frame. One of them is $V=a\phi^2+b\phi^3+c\phi^4$, with constant $a$, $b$, $c$. The other is of the form $\lambda(\phi)\phi^4$, where $\lambda(\phi)$ runs with non-negligible $\log \phi$ and $(\log \phi)^2$ terms. With a choice of suitable parameters, the interplay of the different terms in each potential allows the appearance of  a localized region where ultra slow-roll can be realized. Imposing the requirements mentioned above, this region tends to feature a local minimum, not only in these concrete models but also in many others that have been proposed with similar properties. Satisfying all the requirements with these potential shapes is not easy. Indeed, the potentials of~\cite{Ballesteros:2017fsr} tend to have a scalar spectral index that is smaller than the one preferred by CMB data, once the other conditions are met. There are however ways to fix this problem, e.g.~modifying the potentials slightly or considering minimal extensions of the $\Lambda$CDM model of cosmology~\cite{Ballesteros:2020qam}. 

Ultra slow-roll inflation is the most studied mechanism that has been considered to produce PBH from inflation. The examples discussed above illustrate that realizing the idea of PBH dark matter from inflation in detail requires a level of complexity and tuning that departs from the most minimal models of inflation compatible with the current cosmological data. 

\section{Breakdown of perturbation theory?}

In 2022 it was argued that perturbation theory breaks at CMB scales in models of inflation featuring an ultra slow-roll phase leading to a large spectrum of curvature fluctuations at small scales. More precisely, it was argued that $\mathcal{P_\zeta}$ has to be much smaller than $10^{-2}$ for perturbation theory to hold at CMB scales~\cite{Kristiano:2022maq}. If correct, this result hints at an impossibility to make consistent predictions --at least with currently known methods-- in the regime of perturbation theory that is widely thought to be required to account for all the dark matter with PBH. Interestingly, it also implies that quantum loops of large momenta can have a substantial impact at very large (distance) scales. To describe in a simple way dynamics like the one of the models discussed above, reference \cite{Kristiano:2022maq} assumed a simple piece-wise parametrization of $\eta = \dot\epsilon/(\epsilon H)$ (where $\epsilon = -\dot H/ H^2\ll 1$) with abrupt transitions between constant values of $\eta$ describing slow-roll and ultra slow-roll. It then considered a specific cubic interaction: $m_p^2\, \epsilon\,\dot\eta\,\dot\zeta\,\zeta^2$ and estimated the size of the one-loop correction to $\mathcal{P_\zeta}$ in the limit of zero comoving momentum ($k \rightarrow 0$). Comparing their estimate to the tree-level result lead the authors of \cite{Kristiano:2022maq} to the conclusion that perturbation theory only holds if $\mathcal{P_\zeta}\ll 10^{-2}$. This argument ignited an interesting debate about the validity of perturbation theory in ultra slow-roll models of inflation with a  large tree-level $\mathcal{P_\zeta}$
%~\footnote{The first version of~\cite{Kristiano:2022maq} concluded that PBH formation from single-field inflation had to be ditched altogether.} 
and, also, about the possible effects of small distance scales on very large ones during inflation, see e.g.\ \cite{Inomata:2024lud}.

The problem of the validity of perturbation theory was studied in \cite{Ballesteros:2024zdp} using a gauge in which $\zeta =0$, encoding the primordial fluctuations in the inflaton field. This choice helps to identify all the relevant cubic and quartic interactions to compute the one-loop contribution to $\mathcal{P_\zeta}$, as well as the necessary counterterms to renormalize the ultraviolet divergences. In that reference, the one-loop contribution to $\mathcal{P_\zeta}$ was studied not only in the limit $k \rightarrow 0$ but for all $k$. Instead of the model of~\cite{Kristiano:2022maq}, reference~\cite{Ballesteros:2024zdp} used another one (based on a combination of slow-roll parameters), better suited to describe the interactions in the gauge in which $\zeta =0$. In addition, the possibility of smooth transitions between slow-roll and ultra slow-roll was included using two-parameters: the duration of the ultra slow-roll phase and that of the transitions from and to slow-roll. The computation in~ \cite{Ballesteros:2024zdp} shows that these two parameters determine whether the one-loop contribution to $\mathcal{P_\zeta}$ becomes comparable to its tree-level counterpart at the maximum of the latter. The conclusion reached in~\cite{Ballesteros:2024zdp} is that perturbation theory does not necessarily break in USR models of inflation for PBH dark matter. The analysis of \cite{Ballesteros:2024zdp} also shows that the tree-level, one-loop and counterterm contributions to $\mathcal{P_\zeta}$ have the same dependence on $k$ in the limit $k \rightarrow 0$, becoming indistinguishable. This is one of the reasons why reference \cite{Ballesteros:2024zdp} emphasizes the importance of considering the wave numbers at which $\mathcal{P_\zeta}$ is maximum in order to analyze the validity of perturbation theory. 

\section{Other inflationary mechanisms}

As it was mentioned earlier, other possibilities (different from ultra slow-roll) have been proposed for PBH of inflationary origin. Two will be briefly mentioned here. The first one stems from the convenience of switching the focus to the dynamics of the primordial perturbations, rather than considering the dynamics of the (spatially homogeneous) background inflaton field. This is the tenet of the effective field theory of inflation \cite{Cheung:2007st}. \footnote{See \cite{Weinberg:2008hq} for a different approach to constructing an effective theory of inflation.}  Considering the most general quadratic action for the curvature fluctuation $\zeta$, one realizes that there are quantities different from $\eta$ (or $\epsilon$) that are susceptible of varying rapidly in time during inflation, producing an enhancement of $\mathcal{P_\zeta}$. This possibility was considered in \cite{Ballesteros:2018wlw,Ballesteros:2021fsp}. An example is the speed of sound of $\zeta$, \footnote{Another is a time-varying ``Planck mass'' and others are combinations of these two and $\epsilon$ and $\eta$.} which can be different from 1 (and time-dependent) in models with a non-canonical kinetic term. Actions for $\zeta$ featuring specific symmetries can help to maintain the validity of perturbation theory in this kind of models. There is still much  to explore in this direction.   

Another way in which a large $\mathcal{P_\zeta}$ may be realized exploits the possibility of the inflaton being substantially coupled to other fields during inflation, in such a way that a transient dissipative phase arises \cite{Ballesteros:2022hjk}. This can happen if the inflaton effectively interacts with particles forming a thermal bath that does not get too swiftly depleted as inflation proceeds. In this kind of situation, the enhancement of the power spectrum of curvature fluctuations can come from a stochastic term of thermal origin that depends on the temperature of the bath and on the strength of the interaction between the inflaton and the bath. The equations governing this system are analogous to the ones in warm inflation models and the search for concrete ways to implement this idea for PBH  dark matter is still in its early infancy. \footnote{See also \cite{Bastero-Gil:2021fac}.}

\section{Gravitational wave signatures}

There is an interesting way in which gravitational waves can provide indirect evidence for PBH dark matter of inflationary origin in the asteroid-mass window. The large fluctuations of $\zeta$ that are required to form an abundant population of PBH act as a source of second-order (induced) gravitational waves, as a direct consequence of the non-linearity of General Relativity. This effect leads to a stochastic background of gravitational waves whose energy density scales as the square of $\mathcal{P_\zeta}$. The mean frequency, $f$, of these waves is approximately related to the mean mass, $M$, of the PBH via $f\simeq 0.4\,(10^{-16} {\rm M}_\odot/M)^{1/2}$~Hz. A stochastic background related to the formation of asteroid-mass PBH from the collapse of dense Hubble patches during radiation domination has a frequency of interest for LISA \cite{Saito:2008jc}. It also has a large enough amplitude to be potentially detectable if all the dark matter is composed by PBH. Interestingly, multiple recent pulsar timing array analyses indicate the possibility of a detection of stochastic background of gravitational waves at nHz frequencies. Among the many ideas that have been proposed as potential explanations are second-order gravitational waves sourced by scalar fluctuations \cite{NANOGrav:2023hvm}. These may correspond to black holes of mass $\sim 0.1$M$_\odot$, which might account for a small fraction of the dark matter only. Inflationary PBH whose associated stochastic background of gravitational waves peaks at LIGO frequencies would have evaporated by now and are strongly constrained by the primordial abundance of light nuclei, see \cite{Ballesteros:2020qam} for a discussion. 

In principle, another possibility to probe the abundance of PBH in the asteroid-mass window is offered by the gravitational waves emitted by their hypothetical mergers \cite{Franciolini:2022htd}. The very high frequency and the smallness of the expected characteristic strains pose a very significant technological challenge, see~ \cite{Aggarwal:2025noe} for a review. Remarkably, there appear to be no proposed astrophysical sources of gravitational waves at MHz-GHz frequencies. 

\section{Summary}

PBH in the asteroid-mass range are a contender to explain the dark matter of the Universe. Single-field inflation can be tweaked to provide a possible origin for PBH dark matter. Gravitational waves can be expected to become a key tool to explore these ideas. Indeed, they already are a way to search for heavier PBH. 

\section*{Acknowledgments}

Funding from the grants PID2021-124704NB-I00 by MCIN/AEI/10.13039/501100011033 and by ERDF A way of making Europe, CNS2022-135613 by MICIU/AEI/10.13039/501100011033 and by the European Union NextGenerationEU/PRTR, and Centro de Excelencia Severo Ochoa CEX2020-001007-S by MCIN/AEI/10.13039/501100011033.

\end{document}